\documentclass[11pt,reqno]{amsart}

\usepackage[left=3cm,top=2.5cm,right=3cm,bottom=2.5cm]{geometry}

\usepackage[english]{babel}    

\usepackage[dvips]{graphics}
\usepackage{subfigure}
\usepackage{graphicx}
\usepackage{epsfig}
\usepackage{hyperref}
\usepackage{framed}
\usepackage{psfrag}
\usepackage{tikz} 

\usepackage{amssymb}
\usepackage{amsbsy}
\usepackage{epsfig}
\usepackage{tikz}

\usepackage[english]{babel}    

\usepackage{multicol}
\usepackage{blkarray}
\usepackage{tikz}
\usepackage{mathtools}
\usepackage{amsthm}

\usepackage{amsfonts}

\begin{document}

\title[Spectral clustering approach to COVID-19]{A spectral clustering approach for the evolution of the COVID-19 pandemic in the state of Rio Grande do Sul, Brazil
}

\author[L. E. Allem]{Luiz Emilio Allem}\email{emilio.allem@ufrgs.br}
\author[C. Hoppen]{Carlos Hoppen}\email{ choppen@ufrgs.br}
\author[M. M. MARZO]{Matheus Micadei Marzo}\email{matheus.marzo@ufrgs.br}
\author[L. S. SIBEMBERG]{Lucas Siviero Sibemberg}\email{lucas.siviero@ufrgs.br}

\address{     
     Instituto de Matem\'{a}tica e Estat\'{i}stica,
     Universidade Federal do Rio Grande do Sul, 
     Av. Bento Gon\c{c}alves, 9500, 
     91509-900, Porto Alegre, RS, Brasil.
}

\thanks{C. Hoppen acknowledges the support of CNPq~308054/2018-0 and FAPERGS~19/2551-0001727-8. M. Marzo and L. Sibemberg are supported by CAPES. CAPES is Coordena\c{c}\~{a}o de Aperfei\c{c}oamento de Pessoal de N\'{i}vel Superior, CNPq is Conselho Nacional de Desenvolvimento Cient\'{i}fico e Tecnol\'{o}gico, and FAPERGS is Funda\c{c}\~{a}o de Amparo \`{a} Pesquisa do Estado do Rio Grande do Sul.}



\begin{abstract}
The aim of this paper is to analyse the evolution of the COVID-19 pandemic in Rio Grande do Sul by applying graph-theoretical tools, particularly spectral clustering techniques, on weighted graphs defined on the set of 167 municipalities in the state with population 10,000 or more, which are based on data provided by government agencies and other sources. To respond to this outbreak, the state has adopted a system by which pre-determined regions are assigned flags on a weekly basis, and different measures go into effect according to the flag assigned. Our results suggest that considering a flexible approach to the regions themselves might be a useful additional tool to give more leeway to cities with lower incidence rates, while keeping the focus on public safety.  Moreover, simulations show the dampening effect of isolation on the dissemination of the disease.
\end{abstract}

\maketitle

\section{Introduction}

The aim of this paper is to employ graph-theoretical tools to understand the dissemination of COVID-19 in the Brazilian state of Rio Grande do Sul. These tools may be useful sources of additional information for decision making by health and government authorities.  

The year 2020 has been marked by the global outbreak and spread of the virus SARS-CoV-2, which causes the coronavirus disease (COVID-19) in humans ~\cite{refWHO}. In December 2019, several patients with an unknown severe respiratory disease were traced back to a wholesale market in Wuhan, China. Researchers were quick to detect and isolate a novel strain of coronavirus~\cite{ZZWL20}. It was soon discovered that the virus is highly contagious, and that it can be transmitted by infected individuals before they show the first symptoms and even by infected individuals that remain asymptomatic throughout the course of the disease~\cite{WTQ20}. This has led to unprecedented public health measures by the Chinese authorities. A lockdown of Wuhan and 15 other cities in Hubei Province took effect on January 23~\cite{CYRA20}. On January 30, the World Health Organization declared COVID-19 a public health emergency of international concern ~\cite{refWHO}. In the next month, a large number countries implemented measures aiming to prevent a global pandemic, ranging from travel restrictions, contact tracing and social isolation to border closures and lockdowns~\cite{refWHO}. These actions turned out to be unsuccessful in eradicating the disease, and the WHO characterised the outbreak as a pandemic on March 11. Two days later, it assessed that Europe had become the epicentre of the pandemic~\cite{refWHO}. The virus then quickly reached Brazil. 

The first confirmed case of COVID-19 in Brazil dates back to February 26, in the state of S\~{a}o Paulo, and the Brazilian Health Ministry declared a state of nationwide community transmission on March 20~\cite{DOU200320}. At that point, the number of confirmed cases in the state of Rio Grande do Sul was 37~\cite{painelRS}, and the state government had already instated measures aimed at slowing down the spread of the virus, including school closures and a ban on commercial interstate travel~\cite{DOE190320}. In the next month, a large number of restrictions were imposed on activities that were deemed inessential. By the end of April, there were 1466 cases and 51 deaths officially attributed to COVID-19 in the state of Rio Grande do Sul~\cite{painelRS}. At this point, and as of this writing, there was no vaccine or proven effective treatment for patients with severe cases of COVID-19~\cite{refWHO}. Recognizing the seriousness of the health crisis and the social and economic impact of widespread isolation, the state government unveiled a regulatory model for \emph{controlled distancing}~\cite{DOE100520}\footnote{\url{https://distanciamentocontrolado.rs.gov.br/}}, which went into effect on May 11, and we now briefly describe. 

The state has been divided into 20 (pre-determined) regions based on the availability of ICU beds for COVID-19 patients. Every week, each region is assigned one of four possible \emph{flags}, yellow (low risk), orange (medium risk), red (high risk) or black (very high risk), according to a numerical value based on several indices that measure the spread of the disease and the availability of ICU beds. Each flag entails different social distancing measures and imposes different constraints on businesses (or even their mandatory closure). This regulation has legal precedence over more flexible measures determined by local authorities or by the federal government~\cite{DOU010620}. Due to its effect on daily lives and on the economic activity, this model has been in the spotlight, and it has mustered praise, but also faced criticism. We should mention that, after the adoption of this system in Rio Grande do Sul, other states have followed suit and devised similar models (for instance, Acre, Mato Grosso, Mato Grosso do Sul, Par\'{a}, Rio de Janeiro, and São Paulo).  

The aim of this paper is to analyse the evolution of the COVID-19 pandemic in Rio Grande do Sul by applying graph-theoretical tools on data provided by government agencies and other sources. Given the distancing model described above, we believe that clustering techniques are particularly well-suited for this analysis. Spectral clustering techniques are widely used in exploratory data analysis, but we are not aware of applications in connection with epidemiological models. 

The general idea of clustering is to partition a (typically large) data set into (a much smaller number of) clusters in a way that data in a same cluster are \emph{similar}, and data in different clusters are \emph{dissimilar}. Formal measures of affinity and of the quality of a given partition rely heavily on the context of the problem being considered. In this paper, we address clustering from a graph-theoretical perspective. We consider weighted graphs $G=(V,E,\omega)$, where the \emph{vertex set} $V$ is the data set, the edge set $E$ contains edges connecting elements of $V$ and the function $\omega \colon E \rightarrow \mathbb{R}_{>0}$ assigns a positive weight $w_{ij}$ to each edge $ij \in E$. 

Here, we consider two types of affinity measures. The first type is based on pendulum migration between cities, by which we mean the daily flow of commuters for work or education, to which we incorporate data about self-isolation. According to our results, isolation did not have a strong effect on the way cities are clustered together, suggesting that general appeals for isolation by state and federal authorities, and by the media, have a much stronger impact than appeals by local authorities. The second type of affinity measures is based on the availability of ICU beds. In this case, incorporating weekly data, and considering a more flexible approach to the regions, by which new clusters are determined on a weekly basis, seems to generate useful complementary information for the flag system used in Rio Grande do Sul.

We also used a discrete SEIR compartmental model to simulate the spread of the disease and the effect of the social distancing measures that have been implemented, based on the migration and isolation data used for clustering. In contrast to clustering techniques, models of this type are a basic tool in the epidemiological toolbox, both in their discrete and continuous versions, and there is a vast literature related with them, see~\cite{li2001global,linka2020outbreak} and the references therein. Our contribution in this respect was to show that the data for pendulum migration and isolation, combined with the available disease information, have been successful in explaining the evolution of the disease in the state. Extrapolating from this, we conclude that isolation measures have been very important in slowing down the spread of the disease (often referred to as \emph{flattening the curve}). 

The remainder of the paper is organized as follows. In Section~\ref{sec_data}, we describe the data used in this paper. Section~\ref{sec_clustering} is concerned with spectral clustering and its mathematical foundations. The affinity measures mentioned above are discussed in that section, and we also analyse the partitions that have been obtained by spectral methods. The SEIR model is introduced and analysed in Section~\ref{sec_seir}. We finish the paper with concluding remarks.

\section{Data}\label{sec_data}

In this section, we describe the data used in our study. The actual matrices are available as an appendix, included as an ancillary file. We consider the 167 municipalities in the state of Rio Grande do Sul whose estimated population in 2019 is above 10.000 according to the Brazilian Institute of Geography and Statistics (IBGE)~\footnote{\url{https://www.ibge.gov.br/estatisticas/sociais/populacao}}. Hereafter they will be referred to as \emph{cities}. The distance between cities is given by a square matrix $\mathcal{D}=(d_{ij})$ of order $n=167$, where $d_{ij}$ denotes the average road distance from the seat of $i$ to the seat of $j$ and vice-versa, as calculated by the web mapping service Google Maps. 

Using data from the population census of 2010, which is the most recent census performed in Brazil, we define square matrices $\mathcal{T}=(t_{ij})$ and $\mathcal{E}=(e_{ij})$ of order $n$, where $t_{ij}$ is the number of daily commuters who reside in $i$ and work in $j$ and $e_{ij}$ is the number of commuters who reside in $i$ and go to school in $j$. These matrices have been obtained by extracting anonymized census microdata related to long-form questionnaires, which are publicly available\footnote{\url{https://www.ibge.gov.br/estatisticas/sociais/populacao/9662-censo-demografico-2010.html?=&t=downloads}}, and by extrapolating them to the entire city population (adjusted to the 2019 values) using the survey weights that are part of the census microdata. To extract the data from this large dataset, we used a commercial statistical software.  

We also considered data directly related to the spread of the disease, and to the response to it, which has been extracted directly from the state health authorities~\cite{painelRS}, and indirectly through UFRGS websites~\footnote{\url{https://mhbarbian.shinyapps.io/covid19_rs} and \url{https://www.ufrgs.br/coronavis/}}.  

In our approach, the \emph{time} $t$ is measured in weeks, where our weeks correspond to the state's \emph{epidemiological weeks}, which go from Saturday to Friday. Regarding epidemiological data, we consider $N=17$ weeks starting at the week of March 7-13, when the first cases of COVID-19 were officially confirmed in the state, until July 3. We note that most pandemic related data is actually released on a daily basis, but contains fluctuations that may be attributed to administrative procedures. For instance, the number of reported cases and deaths regularly goes down on weekends and holidays, and surges in the first business days thereafter, which suggests that it does not reflect the actual behavior of the disease. Regarding cases and deaths, the weekly data that we collect is simply the overall number of reported cases in a week. Regarding self-isolation and ICU beds occupancy rates, we take the average over the time period. We should point out that the number of ICU beds in the state expanded considerably during the weeks considered, so that the number of total ICU beds in each city is also tracked on a weekly basis. Finally, we point out that these data are only used for $N=8$ weeks, starting at the week between May 2 and 8 (when the model for controlled distancing was unveiled).  
 
The information about self-isolation in each city $i \in [n]=\{1,\ldots,n\}$ is given by values $\beta_i(t) \in [0,1]$ for all $t \in \{1,2,\ldots,N\}$. This is an index developed by In Loco\footnote{\url{https://www.inloco.com.br/}}, a technology firm with offices in Brazil and in the United States, calculated from granular anonymized geolocation data from more than 60 million mobile devices across Brazil. It is defined as the proportion of devices in a city $i$ that stayed within a radius of 450 meters from their habitual home during day $t$~\cite{ACM20,PMPQ20}.

\section{Clustering}\label{sec_clustering}

Consider a set of points $M=\{p_1,\ldots,p_n\}$ such that a weight $w_{ij}\geq 0$ is assigned to each pair of points $p_{i}$ and $p_{j}$, where $i\neq j$ and $i, j\in [n]$. The aim of data clustering is to partition this set of points into classes such that elements of the same class are more alike, while elements of different classes are less alike. The weight $w_{ij}$ measures \emph{affinity} or \emph{similarity} in this context\footnote{We use the word affinity because, in some of our examples, cities that are more different in some aspects will have more affinity to each other.}; the larger the value, the larger their affinity. For general data sets, a large number of similarity measures appear in the literature, and their quality depends on the context in which they are used~\cite{VONL07}. 

Here, points are cities and weights are used to measure whether cities are highly interconnected or not. Several such measures will be considered here. For instance, a simple way to measure interconnection between cities is by simply considering the number of people who commute between them. This leads to the following matrix, where the weight $\alpha_{ij}$ between cities $i$ and $j$ is defined through the matrices $\mathcal{T}$ and $\mathcal{E}$ defined in Section~\ref{sec_data}:
\begin{equation}\label{def_A0}
A_0=(\alpha_{ij}), \textrm{ where }\alpha_{ij}=t_{ij}+t_{ji}+e_{ij}+e_{ji}.
\end{equation}
This choice is justified because, in the context of affinity measures, it is natural to consider symmetric weights.

In order to understand how the interconnection between cities was affected during the pandemic, we also considered weights given by matrices $A(t)$, for $t \in \{1,\ldots,N\}$. To incorporate self-isolation data, we first adjust the rate of self-isolation in each city $i$ in terms of the average isolation $\overline{\beta}_i$, which was calculated using the same cell-phone data for $i$ in the entire month of February, before the implementation of measures to contain the dissemination of COVID-19. We define
\begin{equation}\label{eq_betalinha}
\beta^\ast_i(t)=\max\left\{\frac{\beta_i(t)-\overline{\beta}_i}{1-\overline{\beta}_i},0\right\},
\end{equation}
so that $\beta^\ast_i(t)=0$ if rates of self-isolation are below average (this actually does not happen in our data set after the first week); otherwise, it is a linear interpolation where 0 corresponds to the average rate and 1 to full isolation. We are now ready to define
\begin{equation}\label{def_At}
A(t)=(a_{ij}), \textrm{ where }a_{ij}=\left(1-\beta^\ast_j(t)\right)(t_{ij}+e_{ij})+\left(1-\beta^\ast_i(t)\right)(t_{ji}+e_{ji}).
\end{equation}
The definition of $A(t)$ reflects our belief that it is conceptually more relevant to consider information about isolation in city $j$ to assess the impact on commuting from $i$ to $j$ than information about isolation in city $i$. On the other hand, we understand that the nature of our isolation index, which estimates the number of individuals who never leave their home, could suggest using indices in city $i$ to limit commutes from $i$ to $j$. This has been tested and would have negligible impact on the results. Moreover, it would have been natural to ignore data related to student mobility as of the third week because all in-person school and university operations had already been suspended by then. However, this turned out to make clustering more unstable, perhaps because entries associated with smaller or more remote cities became too small.

\subsection{Normalized cut}
Before introducing the other affinity measures used in this paper, we first describe the framework of our analysis. We think of the data points as vertices in a graph $G=(V,E)$, where we use $V=[n]$ for simplicity. The weight between $p_i$ and $p_j$ is viewed as a weight $\omega(ij)=w_{ij}$ associated with the edge $ij$ of $G$ (if $w_{ij}=0$, we assume that vertices $i$ and $j$ are not adjacent in $G$). 

In general terms, a clustering problem in $G=(V,E)$ consists of finding a partition $V=V_1 \cup \cdots \cup V_k$ of the vertex set into a pre-determined number $k$ of classes, where the partition optimizes some measure of quality of the partition. There are several such measures proposed in the literature~\cite{VONL07}. In this paper, we work with the the normalized cut introduced by Shi and Malik in~\cite{np}. To define it, some additional notation is needed. Given $U \subset V$, let $\overline{U}=V \setminus U$ be the complement of $U$ with respect to $V$. Moreover, for $S,T \subset V$, let $W(S,T)=\sum_{i \in S, j \in T} w_{ij}$. For a partition $\mathcal{P}=\{V_1,\ldots,V_k\}$ of $V$, let 
\begin{equation}\label{def_ncut}
\mbox{NCut}(\mathcal{P}) ={ \sum_{\ell=1}^{k} \frac{\mbox{Cut}(V_\ell,\overline{V_\ell})}{\mbox{Vol}(V_\ell)}},
\end{equation}
where
$$\mbox{Cut}(\mathcal{P}) = \frac{1}{2} \sum_{\ell=1}^{k}W(V_\ell,\overline{V_\ell}) \textrm{ and }\mbox{Vol}(V_\ell)=\sum_{i \in V_\ell} \sum_{j \in V} w_{ij}.$$
Finding an optimal partition in this context is to find a partition $\mathcal{P}$ of $V$ that minimizes the value of $\mbox{NCut}(\mathcal{P})$. Note that this objective function takes both aims of clustering into account. On the one hand, the only weights that appear on numerators of terms in~(\ref{def_ncut}) are weights of edges whose endpoints lie in distinct classes, so that minimizing the function favors partitions such that vertices in different classes have small weight. On the other hand, the denominator of the term associated with $V_i$ in~(\ref{def_ncut}) counts the weight of each edge with both endpoints in $V_i$ twice, while the other edges incident with $V_i$ are only counted once. So, increasing the weight of internal edges would decrease the value of the cut. Unfortunately, the authors of~\cite{np} showed that the problem of finding such a partition is NP-hard for general graphs (even if $k=2$). 

However, this problem is well-suited for a spectral approach. The following definitions are well known in spectral graph theory. The \emph{weighted adjacency matrix} $W=(w_{ij})$ of a graph $G=(V,E)$ with weight function $\omega$ is defined by $w_{ij}=\omega(ij)$ if $ij \in E$ and $w_{ij}=0$ otherwise.
The degree of a vertex $i \in V$ in $G$ is given by
$d_{i}=\sum_{j=1}^{n}w_{ij}.$
The diagonal matrix with the degrees $d_{1},\ldots,d_{n}$ on the diagonal is called the \emph{degree matrix $D$}. 

At this point, we could simply present the procedure that we use to cluster our data; however, we believe that explaining how it works, and its connection to linear algebra, clarifies our approach. The following computation are performed in detail in~\cite{VONL07}. Given a positive integers $n$ and $k$ and a partition $\mathcal{P}=\{V_1,\ldots,V_k\}$ of the vertex set of a graph $G=(V,E)$ with weight function $\omega$ and no isolated vertices, consider the matrix $X_{\mathcal{P}}\in\mathbb{R}^{n\times k}$ whose columns are the $k$ vectors
$\mathbf{x}^{(\ell)} = ({x_1}^{(\ell)},{x_2}^{(\ell)},\dots , {x_n}^{(\ell)})^{T}$ with coordinates
$$
x_j^{(\ell)} =\left\{
               \begin{array}{cl}
                 \frac{1}{\mbox{Vol}(V_\ell)}  &  \mbox{ if }  j \in V_\ell; \\
                 0        & \hbox{\mbox{ otherwise},}
               \end{array}
             \right.
             $$
for all $\ell \in\{1,\ldots,k\}$ and $j\in\{1,\ldots,n\}$. Using the \emph{Laplacian matrix} $L = D - W$ associated with the weighted graph $G$, it turns out that
$$\mbox{NCut}(\mathcal{P}) = \sum_{\ell=1}^{k}\frac{\mbox{Cut}(V_\ell,\overline{V_\ell})}{\mbox{Vol}(V_\ell)} = \sum_{\ell=1}^{k}{\mathbf{x}^{(\ell)}}^TL\mathbf{x}^{(\ell)} = \mbox{tr}(X_{\mathcal{P}}^TLX_{\mathcal{P}}).$$
Writing $Y_{\mathcal{P}} = D^{-\frac{1}{2}}X_{\mathcal{P}}$ we obtain that 
$$\mbox{NCut}(\mathcal{P}) = \mbox{tr}(Y_\mathcal{P}^T(D^{-\frac{1}{2}}LD^{-\frac{1}{2}})Y_{\mathcal{P}})=\mbox{tr}(Y_\mathcal{P}^T\mathcal{L}Y_{\mathcal{P}}),$$
where $\mathcal{L}=D^{-\frac{1}{2}}LD^{-\frac{1}{2}}$ is the \emph{normalized Laplacian matrix} associated with $G$. Therefore finding an optimal partition in the sense of~\cite{np} is equivalent to finding a partition $\mathcal{P}$ that minimizes
$$\mbox{ncut}_k(G)=\min_{\mathcal{Q}} \mbox{NCut}(\mathcal{Q}) = \min_{\mathcal{Q}} \mbox{tr}(Y_{\mathcal{Q}}^T\mathcal{L}Y_{\mathcal{Q}}),$$
where $\mathcal{Q}$ ranges over all partitions of $V$ into exactly $k$ sets. It is easy to see that $Y_{\mathcal{Q}}^{T}  Y_{\mathcal{Q}} = I$, and by  the Rayleigh-Ritz Theorem~\cite[Theorem 13]{MN99}, we have
\begin{equation}\label{eq_relax}
\min_{Y \in \mathbb{R}^{n\times k}, Y^TY = I} \mbox{tr}(Y^T\mathcal{L}Y) = \lambda_1 + \cdots + \lambda_k, 
\end{equation}
where $0=\lambda_1 \leq \cdots  \leq \lambda_k$ are the $k$ smallest eigenvalues of the symmetric matrix $\mathcal{L}$. Moreover, equality is achieved by matrices $Y$ whose columns are orthogonal unit vectors generated by eigenvectors associated with the eigenvalues $ \lambda_1,\ldots,\lambda_k$. As we have discussed, each partition of $V$ into $k$ parts is associated with a matrix $Y$ as above. However, there are matrices $Y$ that are feasible for~(\ref{eq_relax}), but are not of the form $Y_{\mathcal{Q}}$ for any partition $\mathcal{Q}$. This leads to the the following inequality:
\begin{equation} 
\label{aprox}
\mbox{ncut}_k^{rel}(G)=\min_{Y \in \mathbb{R}^{n\times k}, Y^TY = I} \mbox{tr}(Y^T\mathcal{L}Y)  \leq \mbox{ncut}_k(G). 
\end{equation} 
As in usual LP-relaxations, the left-hand side of the inequality (\ref{aprox}) may be computed efficiently and gives a lower bound on the value of an optimal partition. On the other hand, there is no obvious connection between a matrix $Y$ that achieves $\mbox{ncut}_k^{rel}(G)$ in (\ref{aprox}) (i.e. a matrix constructed from eigenvectors associated with the smallest eigenvalues of $\mathcal{L}$) and a partition into $k$ parts $\mathcal{P}$ such that $\mbox{NCut}(\mathcal{P})$ is close to $ \mbox{ncut}_k(G)$. The following heuristic tries to find good quality partitions. To turn the matrix $Y$ into a partition $\mathcal{P}$, it uses a well-known geometric method, known as $K$-means~\cite{Macqueen67somemethods}. One way of assessing the quality of the output partition $\mathcal{P}$ is by looking at the ratio $\mbox{NCut}(\mathcal{P})/\mbox{ncut}_k^{rel}(G) \geq 1$. If this ratio is exactly 1, the partition $\mathcal{P}$ is optimal. Otherwise, it gives an upper bound on the actual value of the ratio $\rho(\mathcal{P})=\mbox{NCut}(\mathcal{P})/\mbox{ncut}_k(G)$ (however, we should mention that the gap between $\mbox{ncut}_k^{rel}(G)$ and $ \mbox{ncut}_k(G)$ may be very large in general).  It is important to mention that this heuristic has been quite successful in practice, we refer to~\cite{NJW01,NLCK05,vonluxburg2008} for more explanation about these empirical findings. Moreover, defining the best choice for the number of clusters $k$ is an important problem with no definitive solution. Parameters that are often used to indicate a good choice of $k$ are the spectral gap (this is the ratio between consecutive eigenvalues, small ratios followed by a larger jump $\lambda_{k+1}/\lambda_k$ indicate that $k$ is a good choice) and the closeness to 0 ($k$ is the number of eigenvalues below a certain threshold), and the stability of the clusters obtained in repeated iterations of the procedure, but other criteria also appear in the literature~\cite{VONL07}. 

We now state the heuristic of Shi and Malik~\cite{np}, iterated $S$ times. Given an affinity matrix $W$ associated with an $n$-vertex graph $G=(V,E)$, do the following:
\begin{enumerate}
\item[(1)] Let $D$ to be the degree matrix associated with $W$ and construct its normalized Laplacian matrix $ \mathcal{L} = D^{-1/2}LD^{-1/2}$, where $L=D-W$ .

\item[(2)] Compute vectors $\mathbf{x}_1,\mathbf{x}_2,\ldots,\mathbf{x}_k \in \mathbb{R}^n$ , where each $\mathbf{x}_i$ is a unit eigenvector associated with the eigenvalue $\lambda_i$, where $\lambda_1,\ldots,\lambda_k$ are the $k$ smallest eigenvectors of $\mathcal{L}$ (counting multiplicity). In the case of repeated eigenvalues, the eigenvectors associated with them must be orthogonal. Form the matrix $X = [\mathbf{x}_1\mathbf{x}_2\dots \mathbf{x}_k] \in \mathbb{R}^{n\times k}$ by stacking these eigenvectors in columns.

\item[(3)] Form the matrix $Y=(y_{ij})$ from $X= (x_{ij})$ by renormalizing each of the rows to have unit length
(i.e. $y_{ij} = x_{ij}/\sum_{j=1}^{n}x_{ij}$).

\item[(4)] for $s=1,\ldots, S$ do (let $\mathcal{Q}$ denote the best partition obtained up to a given step, where the starting partition is arbitrary.)
\begin{enumerate}
\item[(4.1)] Treating the $i$th row of $Y$ as a point $\mathbf{y}_i \in \mathbb{R}^{k}$, split $\{\mathbf{y}_1,\ldots,\mathbf{y}_n\}$ into $k$ clusters $S_1,\ldots,S_k$ via $K$-means.
\item[(4.2)] Let $\mathcal{P}$ be the partition such that vertex $i$ is assigned to cluster $\ell$ if and only if $\mathbf{y}_i$ lies in $S_\ell$.
\item[(4.3)] If $\mbox{NCut}(\mathcal{P})<\mbox{NCut}(\mathcal{Q})$, redefine $\mathcal{Q}$ as $\mathcal{P}$.
\end{enumerate}
			
\item[(5)] Return $\mathcal{Q}$, the partition with minimum Ncut obtained in step (4).
\end{enumerate}

\begin{figure}
\begin{center}
\includegraphics[scale=0.4]{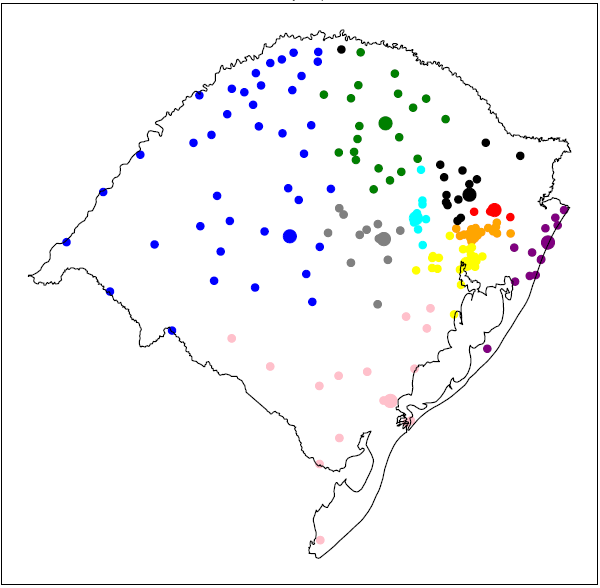}
\caption{Clustering obtained by spectral clustering with respect to measure $A_0$ for $k=10$ clusters. The largest city in each cluster is marked with a larger circle.}
\label{figure_A0}
\end{center}
\end{figure}

When we compute the eigenvalues of the matrix $\mathcal{L}$ associated with the affinity measure $A_0$ defined in~(\ref{def_A0}), we find determine that there is a considerable eigenvalue gap between $\lambda_{10}$ and $\lambda_{11}$, which suggests that $k=10$ is a good choice for the number of clusters. When we apply the above procedure to the affinity measure $A_0$ for $S=500$, we obtain the partition given in Figure~\ref{figure_A0}, whose gap is $\mbox{NCut}(\mathcal{P})/\mbox{ncut}_k^{rel}(G) \approx 1.3256$. This means that $\mathcal{P}$ is at most $32.56 \%$ above the actual value of $\mbox{ncut}_k(G)$, but the gap is typically much smaller (and may possibly be optimal). Regarding stability, this partition $\mathcal{P}$ has been obtained 183 times out of the 500 iterations of the procedure.

\begin{figure}[h]
\begin{center}
\includegraphics[scale=0.3]{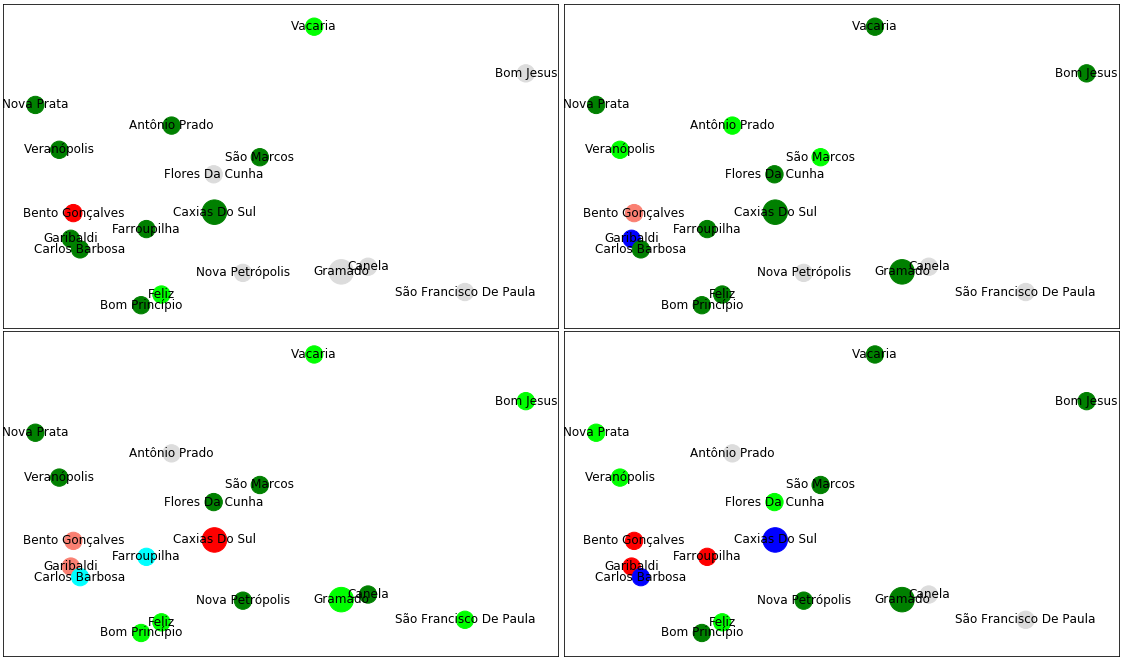}
\caption{Clockwise, starting from the top left. Cases on the weeks from May 2-8, May 9-15, May 16-22 and from May 30 to June 5. Gray stands for no active cases, green for cases in the interval $[1,50]$, blue for $[51,100]$ and red above $100$. Dark colors mean that the number of active cases has increased from the previous week, light colors mean that they have decreased.}
\label{figura_2_clusters}
\end{center}
\end{figure}

Even though the data used to obtain this partition is not related to the pandemic, if we look at the evolution of the number of cases during this time period, the connection between the clusters and the spread of the disease is perceptible.  For instance, Figure~\ref{figura_2_clusters} shows how real data about the disease evolved in cities of two neighboring clusters, marked with black and red in the figure, in four different weeks (material for all regions is available in the ancillary files). The red cluster consists of four cities: Gramado, Canela, Nova Petr\'{o}polis and S\~{a}o Francisco de Paula (which are part of a nationally renowned touristic area) and the other cluster is centered in Caxias do Sul, the second largest city in the state by population. The first cases appear in the cluster of Caxias do Sul quite early, and they quickly spread to cities in the same cluster, which has a relatively large number of active cases by May 2, the first week displayed in the figure (and the ninth week with cases in the state). On the other hand, there are no recorded cases in the cluster of Gramado until the week of May 9. After the first case is identified, all the other cities in the cluster record cases in a span of three weeks. 
This behavior supports the choice of pendulum migration as a footprint for the spread of the disease, as was done in~\cite{SPN20}, for instance. However, instead of census data, the authors of~\cite{SPN20} used mobile geolocation data from~\cite{PMPQ20} to monitor the movement between cities. Also regarding the cities in Figure~\ref{figura_2_clusters}, we point out that they are all part of the same region, according to the state 20 pre-determined regions. 

As mentioned at the beginning of this section, instead of using $A_0$, one could adjust the measure to incorporate rates of isolation, using a different measure $A(t)$ (defined in (\ref{def_At})) at each time $t$. As it turns out, the difference in the partition obtained when performing the above clustering procedure for $A(t)$ instead of $A_0$ is minor. Indeed, the Hamming distance at any time $t$ between the two partitions was at most 1 (out of 167). This may indicate that public response to self-isolation has been rather uniform throughout the state. Calls for self-isolation by state and federal authorities, and by the media, seem to have a much stronger impact than appeals by local authorities.

\subsection{Affinity based on available ICU beds} As mentioned in the introduction, the state government introduced regulation to define when mandatory protocols of social distancing must be put into effect. Every Saturday, each region, out of a pre-determined set of 20 regions (which in turn are sorted into seven \emph{macroregions}), is assigned one of four possible flags, yellow, orange, red or black, according to a numerical value based on several indices, which take the number of cases, the number of hospitalizations, the number of deaths and the availability of ICU beds into account. Once a flag has been assigned, cities in the region must adapt to the state regulations associated with that flag (local governments may enforce stricter rules, if desired).

Even though this method was met by a very positive reception from health and local authorities, its implementation quickly led to complaints by cities and economic agents who deem to have been treated unfairly. For instance, in the first weeks using this method, it was pointed out that several cities where no cases had ever been recorded had been assigned orange or red flags (owing to an outbreak or a shortage of ICU beds in their region, for instance). Moreover, since the index for a region incorporates data from the macroregion to which it belongs, a high risk flag can be assigned to a region in which no city had a substantial number of cases. In some instances, this has led to loud public outcry and threats of disobedience by local authorities, which in turn led to negotiations and adjustments. At the present moment, regulations include automatic `flag reductions' for cities that meet certain criteria. This is the case for cities where no new cases have been recorded in the past two weeks, for instance. Moreover, each city can appeal to a board after its weekly classification has been revealed. When this happens, the city is allowed to present new data, such as an expansion on the total number of ICU beds.

Given this reality, we aim to look at the partition into regions under a more flexible perspective. To this end, we propose affinity measures that consider the availability of ICU beds (updating it weekly) and consider what happens when we re-organize the regions on a weekly basis. For a city $i$, let $u_i(t)$ be the average total number of ICU beds in $i$ at time $t$, and let $\ell_i(t)$ be the average number of ICU beds that are available (i.e. unoccupied and ready to accommodate new patients) in $i$ at time $t$. The first measure is `static', as it only considers the total number of ICU beds at the beginning of the recording process:
\begin{equation}\label{def_C0}
C_0=(\gamma_{ij}), \textrm{ where }\gamma_{ij}=\frac{|u_i(0)-u_j(0)|}{d_{ij}+c},
\end{equation}
where $u_i(0)$ denotes the total number of ICU beds in city $i$ on May 2 and $d_{ij}$ is the distance between $i$ and $j$ given by matrix $\mathcal{D}$ (see Section~\ref{sec_data}) and the constant $c=10$ avoids the effect of very small distances. 

The intuition behind this definition is that the health systems of two cities $i$ and $j$ that are geographically close, but whose health infrastructure is very different, would tend to be interconnected (with the city with small health capability transferring patients to the other), while two cities whose health capacities are equivalent would be less dependent on each other.

The second measure is `dynamic', not only updating the number of ICU beds, but also considering the actual number of ICU beds that are ready to accommodate new patients:
\begin{equation}\label{def_Ct}
C(t)=(c_{ij}(t)), \textrm{ where }c_{ij}=\max\left\{ \eta_i(t) \frac{\ell_j(t)-\ell_i(t)}{d_{ij}+c},  \eta_j(t) \frac{\ell_i(t)-\ell_j(t)}{d_{ij}+c}\right\},
\end{equation}
where $c=10$ and $\eta_i(t)=\frac{u_{i}(t)-\ell_{i}(t)+1}{u_i(t)+1}$. This quantity $\eta_i(t)$ may be viewed as a rate of urgency for city $i$ to look for ICU beds outside its borders. This rate is 1 if it does not have any ICU beds or if all its ICU beds are occupied, and decreases as the percentage of available beds gets larger. The term $\ell_i(t)-\ell_j(t)$ accounts for the fact that a city $j$ with more available ICU beds than $i$ would be desirable to receive patients rom $i$. In other words, the affinity measure of interconnection between $i$ and $j$ goes up from the perspective of $i$ if its health system is strained and $j$ is geographically close and has more available beds. 

\begin{figure}
\begin{center}
\includegraphics[scale=0.3]{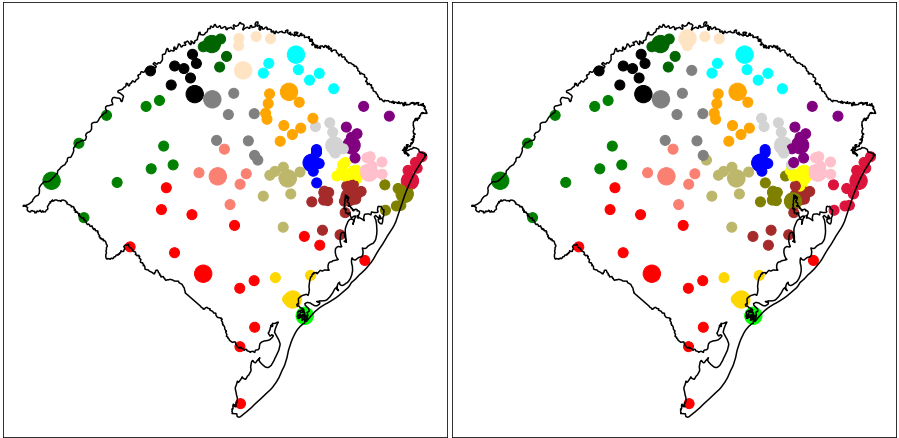}
\caption{Partitions obtained using the affinity measure~(\ref{def_Ct}) using data from the weeks from June 13-19 (left) and June 20-26 (right).}
\label{figura_cluster_dinamico}
\end{center}
\end{figure}

Applying the above spectral partitioning procedure with the affinity measure defined in~(\ref{def_Ct}) for $k=20$ (the number chosen by the state) and $S=500$ produces the partitions in Figure~\ref{figura_cluster_dinamico} in two consecutive weeks. In this particular case, 26 cities switched regions from one week to the next.  

Our aim using this measure is to assess whether allowing the regions to be re-organized on a weekly basis can bring meaningful additional information to one of the features of the state flag system, namely that the state consists of 20 pre-determined regions, which are in turn combined into seven macroregions. To this end, we shall first give a general description of the way in which the state assigns flags to regions, (available at \url{https://distanciamentocontrolado.rs.gov.br/},  a spreadsheet is also available in the ancillary files). The flag is based on 11 individual indices, classified in two main types, disease propagation or healthcare capacity, and computed in one of three levels (within each region, within each macroregion or statewide). For each index, four intervals have been defined, and a flag is assigned to the index according to the interval it belongs to. The flag actually assigned to the region is obtained from a weighted average of the flags assigned to the different indices. 

\begin{figure}[h]
\begin{center}
\includegraphics[scale=0.25]{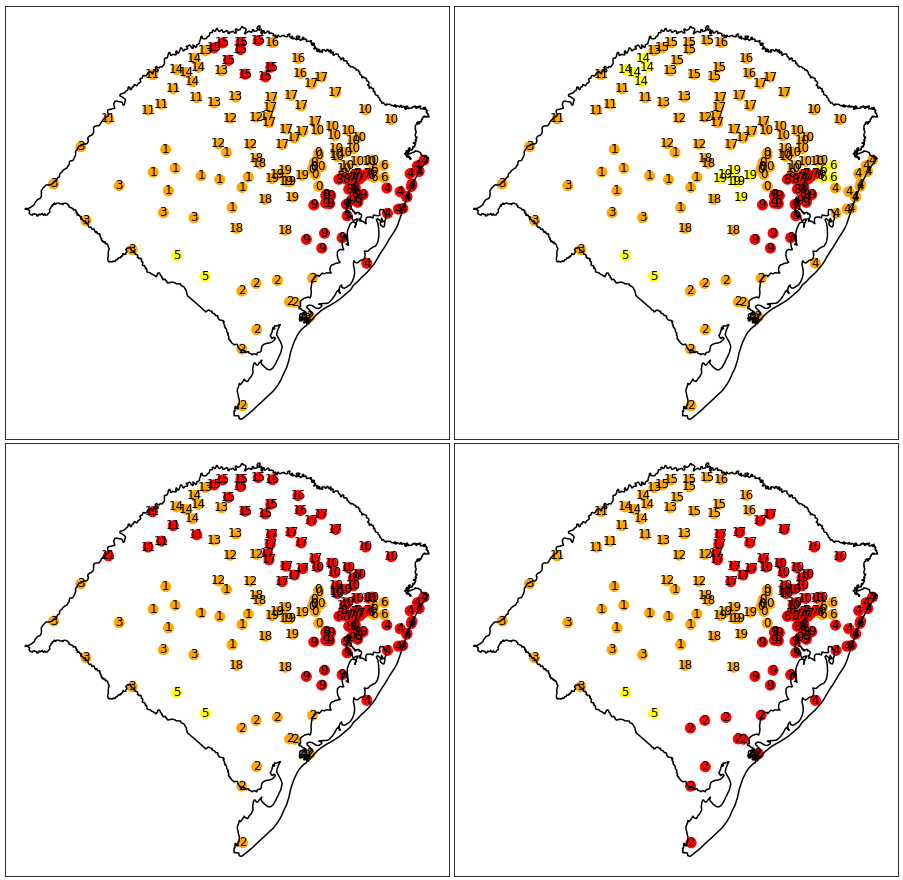}
\caption{Flags assigned by the state formula (left) and by our formula (right) on the weeks from June 13-19 (top) and June 20-26 (bottom).}
\label{figura_compara}
\end{center}
\end{figure}

Here, we have devised an alternative formula (available in the spreadsheet in the ancillary files), which uses exactly the same indices wherever possible. An important difference is that we do not use any indices related with macroregions, as it would not make sense to assign a city to a new region every week, while at the same time assume that cities lie in a fixed macroregion. Unfortunately, some of the data available for macroregions was not publicly available, or was less reliable, for the cities themselves. Because of this, we transferred the weight of these indices to other indices measuring similar features for cities. To assess what the dynamic clustering obtained using our matrices might say about the clustering defined by the state, we proceed in two steps. The first compares the flags assigned to the 20 pre-determined regions using the state's formula and this new formula. Figure~\ref{figura_compara} does this for the weeks from June 13-19 and June 20-26. (A comparison for all seven weeks under consideration may be found in the ancillary files). 
 
\begin{figure}[h]
\begin{center}
\includegraphics[scale=0.25]{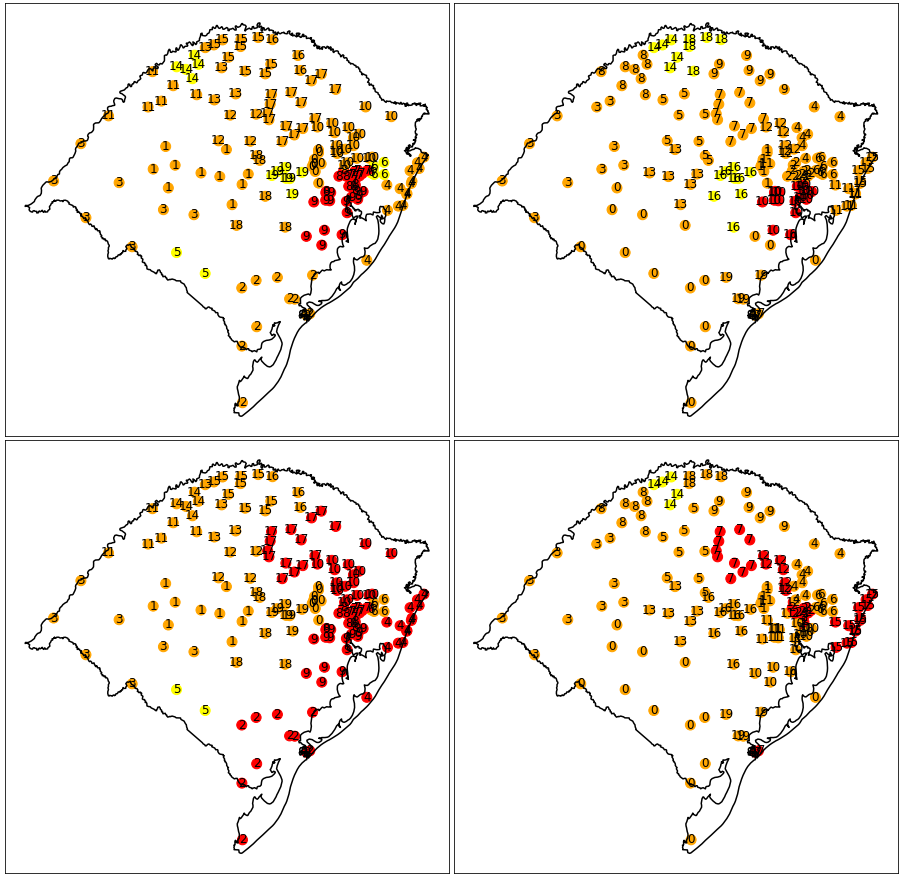}
\caption{Flags assigned by our formula to the state's regions (left) and to the dynamic regions (right) on the weeks from June 13-19 (top) and June 20-26 (bottom). Flags are given by colors, regions are labelled 0 to 19.}
\label{figura_compara2}
\end{center}
\end{figure}

The second step is to split the state in 20 regions on a weekly basis (which we call the \emph{dynamic partition}) and compare the flags assigned by the new formula to these regions and to the 20 pre-determined regions. This is done in Figure~\ref{figura_compara2}, which suggests that more cities that can be assigned a lower-risk flag in the dynamic partition. On the week from June 13-19, 26 cities had a lower-risk flag for the dynamic regions, 13 cities had a higher-risk flag for the dynamic region, and 128 cities remained the same. On the week from June 20-26, these numbers where 50, 2 and 115, respectively.

In short, our computations suggest that the flags assigned with the new formula are related with the flags from the original formula. Moreover, flags assigned in the second step show that partitioning on a weekly basis allows for more flexibility than considering the same partition throughout. This sends the message that it might be possible to devise a formula that takes more, or more reliable, information into account (as in the government's formula), and that allows regions to be adapted on a weekly basis.  

We should emphasize that we do not believe that the new formula presented here is better than the formula used by the state government, quite the opposite, but simulations suggest that the new formula was able to capture the main features of the government's formula using the data available to us. We are also not suggesting that our regions are better than the pre-determined regions defined by the state government. The government's regions are heavily based on the way in which the public health system is organized and on the reality that many cities of small and average size do not have hospitals, particularly hospitals equipped with ICU beds to treat complex cases, and therefore need to establish formal agreements with one or more cities to which their patients can be transferred. Because of this, periodic changes to the regions would require that some cities direct their patients to hospitals in different cities every week, which is certainly not easy to implement. However, in exceptional situations such as a pandemic, this might be justified, and accepted by local governments, given the benefit of more leeway to cities that are not as directly affected by the disease.

\section{SEIR model} \label{sec_seir}

Using the data collected in the previous sections, it is possible define a discrete model for the spread of the disease, which gives a qualitative description of the evolution of the disease and helps us understand the effect of different parameters associated with the disease and of measures to contain it. We consider a discrete susceptible-exposed-infectious-recovered (SEIR) epidemiological model, where the spread of the disease is represented by a recurrence relation indexed by a discrete parameter $t \in \{0,1,\ldots\}$. This recurrence relations are give the expected behavior of a stochastic process defined on a digraph $G=(V,E,\omega)$, where each vertex represents a city and the weight $w_{ij}$ of an arc $ij$ represents the number of commuters from $i$ to $j$ on an average day. Each city $i \in V$ has population $P_i$ and, for all $t \geq 0$, the vector $\mathbf{x}_i(t)=(S_i(t),E_i(t),I_i(t),R_i(t))$ stands for the number of \emph{susceptible}, \emph{exposed}, \emph{infected} and \emph{removed} inhabitants of city $i$ at time $t$, respectively. As usual, all susceptible individuals are assumed to be prone to contracting the disease. Exposed individuals have been infected, but are not yet contagious, while infected individuals are capable of infecting susceptible individuals. Removed individuals either recovered (and became immune from the disease) or passed away. Initially, each city $i$ is assigned a vector $\mathbf{x}_i(0)$ with the number of individuals in each class at the start of the process. 

We now describe how our system evolves. As in the work of Silva, Pereira and Nonato~\cite{SPN20}, we assume that most of the movement between cities may be attributed to daily commutes. On day $t$, part of the population of each city leaves their city to work or study, and comes back in the evening. This leads to a row stochastic matrix $M=(p_{ij})$ of order $n$, where $n=|V|$. We interpret $p_{ij}$ as the \emph{relative flow} from city $i$ to city $j$, given by $p_{ij}=(t_{ij}+e_{ij})/P_i$, where $t_{ij}$ and $e_{ij}$ come from the matrices $\mathcal{T}$ and $\mathcal{E}$ from Section~\ref{sec_data}. This corresponds to the proportion of the population of $i$ that regularly commutes to $j$. The diagonal entries are given by $p_{ii}=1-\sum_{j \neq i} p_{ij}$.

As a consequence, during the day each city $j$ has an effective population of
$$P_j^{\prime}=\sum_{i \in V}p_{ij} P_i.$$
We shall also assume that all classes of individuals are equally likely to move between cities, so that the effective number of individuals of each class in city $j$ on day $t$ is given by 
\begin{eqnarray*}
S_j^{\prime}(t)&=&\sum_{i \in V}p_{ij}(t) S_i(t),~E_j^{\prime}(t)=\sum_{i \in V}p_{ij}(t) E_i(t),\\
I_j^{\prime}(t)&=&\sum_{i \in V}p_{ij}(t) I_i(t),~R_j^{\prime}(t)=\sum_{i \in V}p_{ij}(t) R_i(t).\\
\end{eqnarray*}
In our model, infections only occur during the day (at the city where each individual spends the day). Each such individual is assumed to meet $L$ other individuals in a normal day. However, assuming that a susceptible individual spends the day at city $j$, the number of actual meetings on day $t$ is assumed to be $L(1-\beta^\ast_j(t))^2$, where $\beta^\ast_j(t)$ is the relative rate of isolation of city $j$ on day $t$, given in~(\ref{eq_betalinha}). This rate has been assumed under the simplifying assumption that the probability that, for a meeting to happen, both participants cannot be under self-isolation, and this would happen with probability $(1-\beta^\ast_j(t))^2$ if the decision to self-isolate were taken by each individual spending the day in city $j$, independently of all others, with probability $\beta^\ast_j(t)$. When an individual is infected, we assume that the disease takes its course in 14 days, following the phases described in guidelines of the Center for Disease Control and Prevention (CDC)~\cite{AFGJ20}. In the first four days~\cite{Ma2020}, incubation occurs, in the next 5 days, infected individuals are contagious~\cite{PKGC20} and, in the final five days, individuals are still convalescent, but do not transmit the disease~\cite{imperial}. While infectious, we assume that the probability that an encounter between a susceptible and an infected individual leads to an infection is given by $\tau(k)$, where $k$ is the number of days since the infected individual became contagious. As in~\cite{PKGC20}, we assume that $\tau(k)$ follows a triangular distribution over the five days, with a peak on the third day. We have $\tau(k)=0$ for $k>5$. The area of the triangle in the definition of this distribution is given by $R_0/L$, to ensure that the basic reproductive number (assuming no isolation) is $R_0=2.4$, following a situation report by the WHO~\cite{refWHO2} (see also~\cite{imperial}). 

The recurrence relations become
\begin{eqnarray*}
S_i(t+1)&=& S_i(t) - R_0 S_i(t) \sum_{j} p_{ij} \frac{\sum_{k=t-4}^{t}  (1-\beta^\ast_j(t))^2 I'^{new}_{j}(k) \tau(k-t+5)}{P'_j(t)}\\ 
I^{new}_i(t+1)&=&R_0 S_i(t) \sum_{j} p_{ij} \frac{\sum_{k=t-4}^{t}  (1-\beta^\ast_j(t))^2 I'^{new}_{j}(k) \tau(k-t+5)}{P'_j(t)}\\
E_i(t+1)&=& E_i(t)+  I^{new}_i( t+1 ) - I^{new}_i( t-2 )\\
I_i(t+1)&=& I_i(t) + I^{new}_i(t-2) - I^{new}_i( t-13)\\
R_i(t+1)&=&R_i(t)+ I^{new}_i( t-13)
\end{eqnarray*}
In the above, for simplicity, we assume that $I^{new}_i(s)=0$ for all $s \leq 0$ and $i \in [n]$. Just to illustrate where these equations come from, we discuss the case where an individual in city $i$ does not contract the disease at time $t+1$ in the case where there is no social distancing. With probability $p_{ij}$, the individual moved to city $j$ on day $t+1$. The probability that an encounter leads to an infection is
$$\frac{R_0}{L} \sum_{k=t-4}^{t}  \tau(k-t+5)  \frac{I'^{new}_{j}(k)}{P'_j(t)},$$
so that the probability that no encounter leads to an infection, given that the individual spends the day in city $j$, is
$$\left(1- \frac{R_0}{L} \sum_{k=t-4}^{t}  \tau(k-t+5)  \frac{I'^{new}_{j}(k)}{P'_j(t)}\right)^L \approx 1-R_0 \sum_{k=t-4}^{t}  \tau(k-t+5)  \frac{I'^{new}_{j}(k)}{P'_j(t)}.$$ 
Since the same holds for each susceptible individual in $i$ and knowing the proportion of susceptible individuals that commute from $i$ to $j$, the first equation in the above system gives the expected number of susceptible individuals at time $t$ that remain susceptible at time $t+1$.

\begin{figure}
\begin{center}
\includegraphics[scale=0.3]{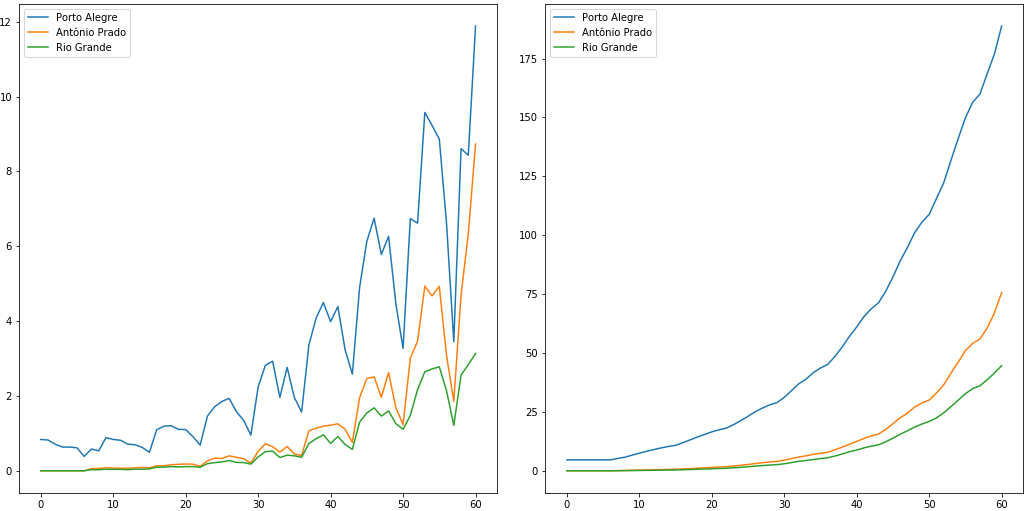}
\caption{Number of new cases and the overall number of cases (per 100,000 inhabitants) in three cities of Rio Grande do Sul. The $x$ axis represents the number of days after May 26.}
\label{figura_seir}
\end{center}
\end{figure}

We run this model starting with the official state data on May 26 to simulate the evolution of the disease until July 9. The number of new infections in the days before this date are estimated using data from May 20-26, where we assume that new cases correspond to 10\% of the number of active cases. The results for the cities of Porto Alegre (the state capital and largest city), Rio Grande (the largest port in Southern Brazil and the city with highest average rate of self-isolation) and Ant\^{o}nio Prado (a small city with a population of about 13,000, where the average rates of self-isolation are lowest) appear in Figure~\ref{figura_seir}. 

\begin{figure}
\begin{center}
\includegraphics[scale=0.25]{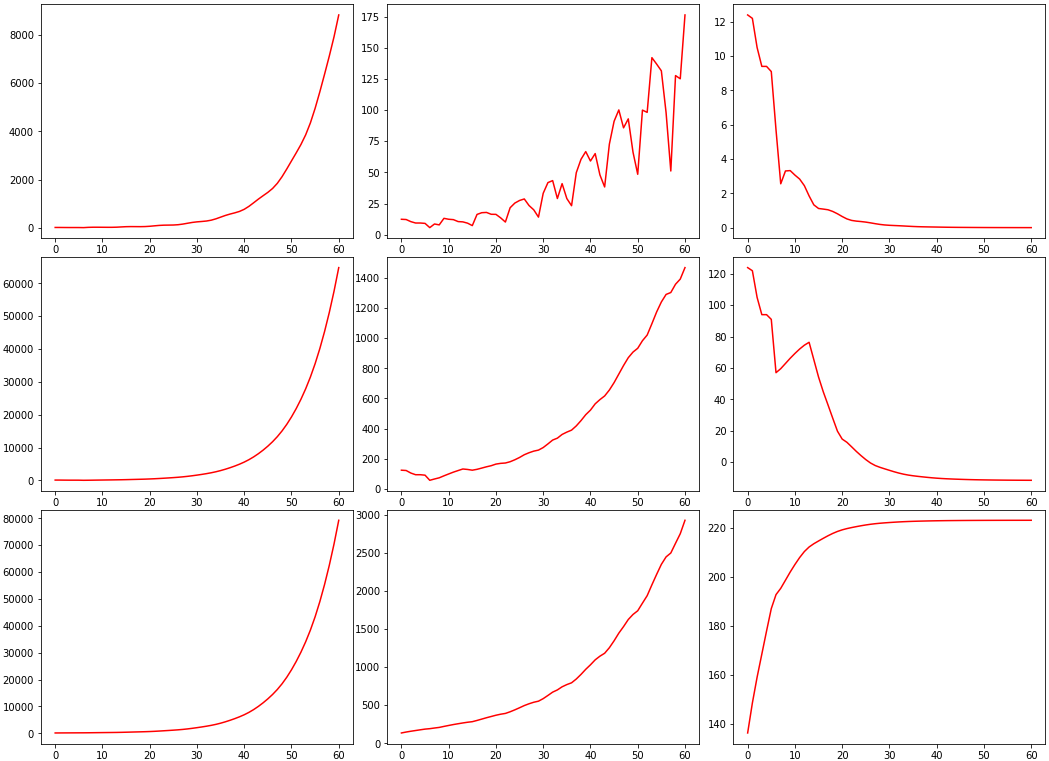}
\caption{On the top: Number of new cases in Porto Alegre assuming the actual isolation data (left), no isolation (center) and strict isolation (right). In the middle: number of active cases in each scenario. On the bottom: overall number of cases in each scenario.}
\label{figura_sem_isolamento}
\end{center}
\end{figure}

It is striking to compare it with the behavior of these quantities in the case where there is no social distancing (that is $\beta^\ast_j(t)=0$ for all $j$ and $t$) and with the situation in which the high rates of self-isolation observed on the week between March 21 and 27 had been maintained after May 26 ($\beta_i=0.614$, on average). This appears in Figure~\ref{figura_sem_isolamento}.

\begin{figure}
\begin{center}
\includegraphics[scale=0.3]{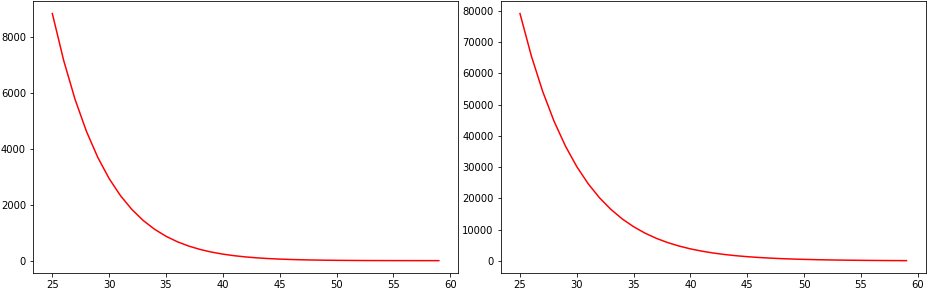}
\caption{Number of active cases, and the overall number of cases (per 100,000 inhabitants) in Porto Alegre on July 9, assuming that the rate of isolation remains constant, and is given by the value on the $x$ axis.}
\label{figura_variacao_isolamento}
\end{center}
\end{figure}

To see the effect of self-isolation in this model, in Figure~\ref{figura_variacao_isolamento} we plot the number of cases in Porto Alegre on July 9 assuming that the rate of self-isolation remained constant throughout the time period, and is given by the corresponding value on the $x$-axis.
According to our data, the average rate of isolation in Porto Alegre has been about 44.3\% during this time period. We note that a simple calculations shows that, while the number of susceptible individuals is much higher than the number of individuals in the other classes, isolation would need to be above 55\% to keep the effective reproductive number of the disease is below 1.

Even though we have opted to plot the evolution of the disease from May 26 to avoid intrinsic errors coming from initial conditions where the number of infected individuals was very small, we should mention that the isolation data were successful at explaining the ups-and-downs in the number of cases in the first weeks of the pandemic in Porto Alegre.  According to the simulations, the number of cases remained stable between March 31 and the week of May 26, and started growing rapidly since then. State data report that the number was stable until early June, and grew rapidly since then. (Specific data are in the ancillary files.) 

\section{Concluding remarks} 

In this paper, we looked at the evolution of the COVID-19 pandemic in Rio Grande do Sul using graph theory. We applied spectral clustering techniques on weighted graphs defined on the set of 167 municipalities in the state with population 10,000 or more, using official data provided by government agencies and isolation data by In Loco. Results related with our first measure, based on pendulum migration, reiterate that pendulum migration is an important means of spreading the disease. Moreover, given the specific context of the flag system in Rio Grande do Sul, our main affinity measure is based on the availability of ICU beds. Our results suggest that considering a flexible approach to the regions themselves might be a useful additional tool in giving more leeway to cities with lower incidence rates, while keeping the focus on public safety. However, this is just a first step in evaluating the adequacy of such an approach. More data (in a municipal level) would be necessary to perform a direct comparison with the government system. Moreover, state and local authorities would need to assess whether periodic changes to the regions could be readily met with changes in patient transfer protocols.

In a different direction, we have observed that disease information from the literature, combined with the isolation data, have provided a coherent qualitative description of the evolution of the pandemic in Rio Grande do Sul using a simple discrete SEIR model. Extrapolating from this, we conclude that isolation measures have been very important in slowing down the spread of the disease. Of course, better results would be achieved with a better understanding of the behavior of the disease and with a model that takes more information into account.

\section*{Acknowledgments}
The authors are particularly indebted to In Loco for providing data about self-isolation in the cities of Rio Grande do Sul and to Prof. M\'{a}rcia Barbian for sharing her data about availability and occupancy of ICU beds in the state. The authors also thank Alisson Matheus Fachini Soares, Guilherme Tadewald Varella and Lucas da Rocha Schwengber for helpful discussions leading to this paper.

\bibliographystyle{ieeetr}
\bibliography{Modelo-TEMA}

\begin{thebibliography}{10}

\bibitem{refWHO}
``{Timeline of WHO's response to COVID-19}.''
  "\url{https://www.who.int/news-room/detail/29-06-2020-covidtimeline}", 2020.

\bibitem{ZZWL20}
N.~Zhu, D.~Zhang, W.~Wang, X.~Li, B.~Yang, J.~Song, X.~Zhao, B.~Huang, W.~Shi,
  R.~Lu, P.~Niu, F.~Zhan, X.~Ma, D.~Wang, W.~Xu, G.~Wu, and G.~Gao, ``{A novel
  coronavirus from patients with pneumonia in China, 2019},'' {\em New England
  Journal of Medicine}, vol.~382, 01 2020.

\bibitem{WTQ20}
Y.~Wang, J.~Tong, Y.~Qin, T.~Xie, J.~Li, J.~Li, J.~Xiang, Y.~Cui, E.~S. Higgs,
  J.~Xiang, and Y.~He, ``{Characterization of an asymptomatic cohort of
  SARS-COV-2 infected individuals outside of Wuhan, China},'' {\em Clinical
  Infectious Diseases}, 05 2020.

\bibitem{CYRA20}
D.~Cyranoski, ``{What China's coronavirus response can teach the rest of the
  world},'' {\em Nature}, vol.~579, no.~7800, pp.~479--480, 2020.

\bibitem{DOU200320}
``{Di\'{a}rio Oficial da Uni\~{a}o, Portaria 454, 20 de mar\c{c}o de 2020,
  Minist\'{e}rio da Sa\'{u}de},'' March 2020.

\bibitem{painelRS}
``{Painel Coronav\'{i}rus RS- Secretaria Estadual de Sa\'{u}de},'' 2020.

\bibitem{DOE190320}
``{Di\'{a}rio Oficial do Estado do Rio Grande do Sul, Decreto 55.128, 19 de
  mar\c{c}o de 2020},'' March 2020.

\bibitem{DOE100520}
``{Di\'{a}rio Oficial do Estado do Rio Grande do Sul, Decreto 55.240, 10 de
  maio de 2020},'' May 2020.

\bibitem{DOU010620}
``{Di\'{a}rio Oficial da Uni\~{a}o, Decis\~{a}o A\c{c}\~{a}o Direta de
  Inconstitucionalidade 6343, 1 de junho de 2020, Supremo Tribunal Federal},''
  June 2020.

\bibitem{li2001global}
M.~Y. Li, H.~L. Smith, and L.~Wang, ``Global dynamics of an seir epidemic model
  with vertical transmission,'' {\em SIAM Journal on Applied Mathematics},
  vol.~62, no.~1, pp.~58--69, 2001.

\bibitem{linka2020outbreak}
K.~Linka, M.~Peirlinck, F.~Sahli~Costabal, and E.~Kuhl, ``Outbreak dynamics of
  covid-19 in europe and the effect of travel restrictions,'' {\em Computer
  Methods in Biomechanics and Biomedical Engineering}, pp.~1--8, 2020.

\bibitem{ACM20}
N.~Ajzenman, T.~Cavalcanti, and D.~Da~Mata, ``More than words: Leaders' speech
  and risky behavior during a pandemic,'' {\em SSRN}, 04 2020.

\bibitem{PMPQ20}
P.~S. Peixoto, D.~R. Marcondes, C.~M. Peixoto, L.~Queiroz, R.~Gouveia,
  A.~Delgado, and S.~M. Oliva, ``Potential dissemination of epidemics based on
  brazilian mobile geolocation data. part i: Population dynamics and future
  spreading of infection in the states of sao paulo and rio de janeiro during
  the pandemic of covid-19.,'' {\em medRxiv}, 2020.

\bibitem{VONL07}
U.~Von~Luxburg, ``A tutorial on spectral clustering,'' {\em Statistics and
  computing}, vol.~17, no.~4, pp.~395--416, 2007.

\bibitem{np}
{Jianbo Shi} and J.~{Malik}, ``Normalized cuts and image segmentation,'' {\em
  IEEE Transactions on Pattern Analysis and Machine Intelligence}, vol.~22,
  no.~8, pp.~888--905, 2000.

\bibitem{MN99}
J.~Magnus and H.~Neudecker, {\em Matrix Differential Calculus with Applications
  in Statistics and Econometrics (Revised Edition)}.
\newblock John Wiley \& Sons Ltd, 1999.

\bibitem{Macqueen67somemethods}
J.~Macqueen, ``Some methods for classification and analysis of multivariate
  observations,'' in {\em In 5-th Berkeley Symposium on Mathematical Statistics
  and Probability}, pp.~281--297, 1967.

\bibitem{NJW01}
A.~Y. Ng, M.~I. Jordan, and Y.~Weiss, ``On spectral clustering: Analysis and an
  algorithm,'' in {\em Proceedings of the 14th International Conference on
  Neural Information Processing Systems: Natural and Synthetic}, (Cambridge,
  MA, USA), pp.~849--856, MIT Press, 2001.

\bibitem{NLCK05}
B.~Nadler, S.~Lafon, R.~R. Coifman, and I.~G. Kevrekidis, ``Diffusion maps,
  spectral clustering and eigenfunctions of fokker-planck operators,'' in {\em
  Proceedings of the 18th International Conference on Neural Information
  Processing Systems}, NIPS'05, (Cambridge, MA, USA), pp.~955--962, MIT Press,
  2005.

\bibitem{vonluxburg2008}
U.~von Luxburg, M.~Belkin, and O.~Bousquet, ``Consistency of spectral
  clustering,'' {\em Ann. Statist.}, vol.~36, pp.~555--586, 04 2008.

\bibitem{SPN20}
P.~J.~S. Silva, T.~Pereira, and L.~G. Nonato, ``Robot dance: a city-wise
  automatic control of covid-19 mitigation levels,'' {\em medRxiv}, 2020.

\bibitem{AFGJ20}
B.~Adhikari, L.~Fischer, B.~Greening, S.~Jeon, E.~Kahn, G.~Kang, G.~Rainisch,
  M.~Meltzer, and M.~Washington, ``{COVID19Surge: a manual to assist state and
  local public health officials and hospital administrators in estimating the
  impact of a novel coronavirus pandemic on hospital surge capacity},'' 2020.

\bibitem{Ma2020}
S.~Ma, J.~Zhang, M.~Zeng, Q.~Yun, W.~Guo, Y.~Zheng, S.~Zhao, M.~H. Wang, and
  Z.~Yang, ``Epidemiological parameters of coronavirus disease 2019: a pooled
  analysis of publicly reported individual data of 1155 cases from seven
  countries,'' {\em medRxiv}, 2020.

\bibitem{PKGC20}
C.~M. Peak, R.~Kahn, Y.~H. Grad, L.~M. Childs, R.~Li, M.~Lipsitch, and C.~O.
  Buckee, ``Comparative impact of individual quarantine vs. active monitoring
  of contacts for the mitigation of covid-19: a modelling study,'' {\em
  medRxiv}, 2020.

\bibitem{imperial}
``{Report 9: Impact of non-pharmaceutical interventions to reduce COVID-19
  mortality and healthcare demand, Imperial College COVID-19 Response Team},''
  2020.

\bibitem{refWHO2}
``{Coronavirus disease 2019 (COVID-19) Situation Report 46, World Health
  Organization},'' 2020.

\end{thebibliography}

\end{document}